# Comment on arXiv:1506.06758
# Terahertz time-domain spectroscopy of transient metallic and superconducting states

D. Nicoletti[1,*], M. Mitrano[1], A. Cantaluppi[1], and A. Cavalleri[1,†]

[1]*Max Planck Institute for the Structure and Dynamics of Matter, Hamburg, Germany*
*daniele.nicoletti@mpsd.mpg.de
†andrea.cavalleri@mpsd.mpg.de

We comment on the model proposed by Orenstein and Dodge in *arXiv:*1506.06758v1, which describes time-domain terahertz measurements of transiently generated, high-electron-mobility (or superconducting) phases of solids. The authors' main conclusion is that time-domain terahertz spectroscopy does not measure a response function that is mathematically identical to the transient optical conductivity. We show that although this is correct, the difference between the measured response function and the microscopic optical conductivity is small for realistic experimental parameters. We also show that for the experiments reported by our group on light-induced superconducting-like phases in cuprates and in organic conductors, the time-domain terahertz yields a very good estimate for the optical conductivity.

Time-resolved THz spectroscopy (THz-TDS) has been used to characterize transient superconducting-like states in photo-stimulated high-$T_c$ cuprates [1-6] and, more recently, in the organic compound $K_3C_{60}$ [7]. These exotic states were typically induced by excitation with femtosecond mid-infrared pulses [1-4,7], which were tuned to specific infrared-active phonon resonances. In some cases, near-infrared light was also used to switch between charge order and superconductivity [5,6].

Prototypical fingerprints of transient superconductivity were identified in the snapshots of the frequency-dependent optical conductivity $\sigma(\omega,t)$, which was extracted from the measured THz-TDS response $\Sigma(\omega,t)$ at different pump-probe time delays $t$. In the $K_3C_{60}$ measurements [7], the real part of the extracted conductivity $\sigma_1(\omega,t)$ exhibited an optical gap, whereas the imaginary part $\sigma_2(\omega,t)$ was found to diverge toward low frequencies. Other features, such as a reflectivity edge reminiscent of the Josephson Plasma Resonance in the cuprates, were also taken as indication of transient superconducting-like behavior.

However, crucial to the interpretation of these measurements is the assumption that the Fourier-transformed THz-TDS trace $\Sigma(\omega,t)$ can, in the parameter range explored by the experiment, yield the transient frequency-dependent conductivity $\sigma(\omega,t)$.

Orenstein and Dodge [8] discuss the extent to which this assignment is reliable. They propose and analyze a model in which $\delta n$ carriers are created by photoexcitation. We note that a model based on photoexcited carriers is in our view not the most appropriate description for the case of vibrational excitation in solids, for which one expects a quench in the Hamiltonian parameters at constant carrier density. Indeed, Orenstein and Dodge do mention the possibility of a quench of Drude momentum relaxation rate as an alternative scenario for these experiments, although one such situation is not discussed quantitatively in their paper.

In the model discussed in Ref. [8], the authors show that the TDS response function $\Sigma(\omega,t)$ is never exactly equivalent to the change in optical conductivity $\delta\sigma(\omega,t)$. This effect, which is especially important near zero pump-probe time delay, has already been discussed in the literature in the context of "perturbed free induction decay" at optical frequencies [9,10] and in other papers analyzing artifacts in THz spectroscopy [11-13].

Here, we argue that even for this "photo-carrier" model, the differences between $\Sigma(\omega,t)$ and $\delta\sigma(\omega,t)$ are pronounced only in *unphysical* limits. To find large deviations, one needs to artificially set infinitely short risetimes and momentum relaxation times far longer than the few picosecond lifetime observed in our experiments.

In Fig. 1, we report selected snapshots of the calculated response function $\Sigma(\omega,t)$ from Eq. (13) in Ref. [8]. Panel 1a displays a typical trace reported in [8]. If one assumes an *infinitely short* excitation time and a transient state with *100 ps long relaxation time* (hence far longer than that determined experimentally) the TDS function $\Sigma(\omega,t)$ (continuous grey curve in 1a) is indeed very different from $\delta\sigma(\omega,t)$ (dashed grey curve in 1a). Large oscillations, whose period evolves with time delay $t$ (not shown here) [8], are expected.

However, simply by allowing for a short relaxation time $T \simeq 1$ ps [1-7], one finds that the oscillations in $\Sigma(\omega,t)$ are strongly reduced (continuous blue curve). For the experimentally measured frequency range (unshaded region), $\Sigma(\omega,t)$ is already very similar to $\delta\sigma(\omega,t)$ (dashed blue curve).

Starting from the blue curves of Fig. 1a, if one then also assumes a realistic risetime for the signal, the matching between $\Sigma(\omega,t)$ and $\delta\sigma(\omega,t)$ improves further. This can be appreciated by comparing continuous ($\Sigma(\omega,t)$) and dashed ($\delta\sigma(\omega,t)$) red curves in Fig. 1b, calculated after replacing the Heaviside Theta functions in Eq. (9) of Ref. [8] with a more appropriate Gauss error function with risetime $\Delta t = 400$ fs [7].

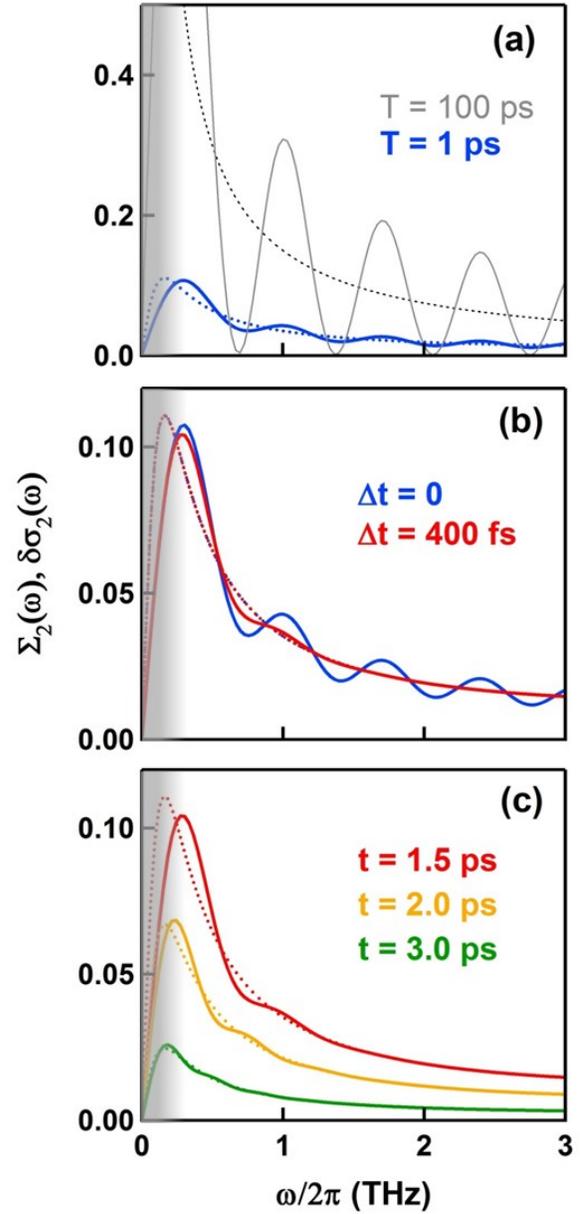

**Fig. 1:** Imaginary part of the TDS response function $\Sigma_2(\omega,t)$ (solid lines) and imaginary conductivity $\delta\sigma_2(\omega,t)$ (dashed lines) as calculated for a Drude model after photoexcitation for different parameter values: $T$ =100 ps, $\Delta t$ = 0, $t$ = 1.5 ps (grey); $T$ =1 ps, $\Delta t$ = 0, $t$ = 1.5 ps (blue); $T$ =1 ps, $\Delta t$ = 400 fs, $t$ = 1.5 ps (red); $T$ =1 ps, $\Delta t$ = 400 fs, $t$ = 2.0 ps (orange); $T$ =1 ps, $\Delta t$ = 400 fs, $t$ = 3.0 ps (green). The frequency axis is expressed in units of $2\pi$ to match the scale of Ref. [8], where a factor of $2\pi$ is missing. All curves in panels b and c are normalized by $\sigma_0 = \delta n(0)Te^2/m$ as in Ref. [8]. The frequency range $\omega \lesssim 1/T$ is shaded in grey.

COMMENT ON ARXIV:1506.06758

The new expression for $\Sigma(\omega, t)$ reads:

$$\Sigma(\omega, t) = C \int_{-\infty}^{+\infty} \left\{ \left[1 + erf\left(\frac{t-\tau}{\Delta t}\right)\right] \left[1 + erf\left(\frac{t}{\Delta t}\right)\right] e^{-t/T} \Theta(\tau) e^{-\tau/T} \right\} e^{i\omega\tau} d\tau$$

where $C$ is a constant prefactor, $\tau$ and $\omega$ are the conjugate dynamical variables, $t$ is the pump-probe time delay, $\Delta t$ is the risetime, and $T$ is the photocarrier lifetime.

Finally, if one probes at longer time delays ($t \geq 2$ ps), for which the experimental TDS signal is still finite, the matching becomes nearly perfect (see orange and green curves in Fig. 1c).

Hence, once realistic parameters are considered, the small discrepancies between $\Sigma_2(\omega, t)$ and $\delta\sigma_2(\omega, t)$ do not influence the qualitative assignment of a superconducting-like phase. A similar analysis, not reported here, can also be obtained for the real part of the conductivity $\delta\sigma_1(\omega, t)$.

In conclusion, starting from the photo-carrier model introduced by Orenstein and Dodge [8], and using the rise and decay times extracted from the experiments of Ref. [1-7], we show here that the measured $\Sigma(\omega, t)$ yields a reliable $\delta\sigma(\omega, t)$.

Future work should address realistic quenches more comprehensively and explore experimental boundaries over which transient states can be determined. Secondly, longer excitation pulses that extend the lifetime of the non-equilibrium state will make it possible to use different probing, including fast electrical transport or magnetic measurements, to reveal the true nature of the transient coherent phases.